\documentclass[aps,twocolumn,showpacs,preprintnumbers,prb]{revtex4} 
 
\usepackage[english]{babel} 
\usepackage{amsmath} 
\usepackage{graphicx}
\usepackage{dcolumn}
\usepackage{bm}
\usepackage{array} 
\usepackage{color}

\begin{document} 
 
\title{Energy loss of charged particles moving parallel to a magnesium surface} 
 
\author{M. G. Vergniory$^{1,2}$, V. M. Silkin$^{2,3}$, 
 I. G. Gurtubay$^{1,2}$, and J. M. Pitarke$^{1,4}$} 
\affiliation{ 
$^1$Materia Kondentsatuaren Fisika Saila, Euskal Herriko Unibertsitatea, 
644 Posta kutxatila, E-48080 Bilbo, Basque Country, Spain\\ 
$^2$Donostia International Physics Center (DIPC),\\ 
Manuel de Lardizabal Pasealekua, E-20018 Donostia, Basque Country, Spain\\ 
$^3$Materialen Fisika Saila, Euskal Herriko Unibertsitatea, 1072 Posta kutxatila, E-20080 Donostia, Basque Country, Spain\\ 
$^4$CIC nanoGUNE Consolider and Centro F\'{\i}sica Materiales 
CSIC-UPV/EHU, Mikeletegi Pasealekua 56, E-20009 Donostia, Basque 
Country, Spain} 
 
\date{\today} 
 
\begin{abstract} 
We present {\it ab initio} calculations of the electronic energy loss of 
charged particles 
 moving outside a magnesium surface, from a realistic description of the 
one-electron band structure and a full treatment of the dynamical electronic 
response of valence electrons. Our results indicate that the finite width of 
the plasmon resonance, which is mainly due to the presence of band-structure 
effects, strongly modifies the asymptotic behaviour of the energy loss at 
large distances from the surface. This effect is relevant for the 
understanding of the interaction between charged particles and the internal 
surface of microcapillaries. 
\end{abstract} 
 
\pacs{71.10.Ca, 71.45.Gm, 73.20.At, 78.47.+p} 
 
\maketitle 
 
\section{Introduction} 
 
Collective excitations at metal surfaces (surface plasmons, 
SP)~\cite{ritchie1} are well known to play a key role in a wide 
spectrum of science, ranging from physics and materials science to 
biology.\cite{review} Here we focus on one specific situation where 
surface plasmons play a key role: the energy loss of charged 
particles moving near a metal surface. This work has been partially 
motivated by recent theoretical and experimental studies of the 
interaction of highly charged ions\cite{arnau} with the internal 
surface of microcapillaries and 
nanocapillaries,~\cite{yamazaki98,sto,tokesi} whose interpretation 
calls for an understanding of the asymptotic behaviour of the energy 
loss at large distances from the surface. Our {\it ab initio} 
calculations of the electronic energy loss of charged particles 
moving outside a magnesium surface indicate that the finite width of 
the SP resonance, which is mainly due to the presence of 
band-structure effects,~\cite{silkin} strongly modifies this 
asymptotic behaviour. 
 
When a charged particle is placed in front of a metal surface, the distribution 
of electrons in the surface and the bulk is modified, an induced charged 
density is built up, and both single-particle and collective excitations 
are created which are ultimately responsible for the electronic energy 
loss of the external charged particle. Moving charged particles can also lose 
energy as a result of the interaction with the nuclei of the solid; however, 
this energy loss is negligible compared to the electronic energy loss, unless 
the probe charge moves at velocities that are extremely small compared to the 
Fermi velocity $v_F$ of the solid. 
 
For a weak interaction between the external charged particle and the electrons 
of the metal surface, the electronic response can be treated within 
linear-response theory. In the case of charged particles moving inside a 
solid, nonlinear effects are known to be crucial in the interpretation of 
energy-loss measurements;~\cite{nazarov1,nazarov2} however, nonlinear 
corrections have been shown to be less important when the charged particle 
moves outside the solid.~\cite{bergara1} On the other hand, the electronic response 
of the metal surface is expected to be strongly dependent upon the actual 
structure of the ground-state electron density. Therefore, an accurate 
description of the electronic energy loss of charged particles moving near a 
metal surface lies mainly in the understanding of two basic ingredients: the 
electronic properties of the ground state of the target and the linear 
response of a many-electron system to external perturbations. 
 
Existing self-consistent calculations of the energy-loss spectra at solid 
surfaces are based on either the jellium model~\cite{aranprb} or a 
one-dimensional (1D) model potential that still assumes translational 
invariance along the surface.~\cite{alducin} Here we report the first 
three-dimensional (3D) {\it ab initio} calculation of the electronic energy 
loss that incorporates the full band structure of the solid surface. 
Specifically, we calculate from first principles the energy loss of charged 
particles that move parallel to the (0001) surface of Mg. We use the random-phase 
approximation (RPA),~\cite{fetter} and we focus on the limit in which the moving particle travels at large 
distances (compared to the Fermi wavelength) from the surface. We demonstrate that even for 
a free-electron-like metal such as Mg band-structure effects on the finite width of the SP 
resonance strongly modify this asymptotic behaviour. 
 
The paper is organized as follows. First of all, in Sec. II we describe our 
full {\it ab initio} treatment of the wavevector- and frequency-dependent electronic response 
of valence electrons, based on a realistic description of the one-electron 
band structure. From the knowledge of the density-response function we derive the 
surface-response function of the system, whose imaginary part is related to the rate at which an 
external potential generates electronic excitations at the solid surface. In Sec. III, we derive an 
explicit expression for the electronic energy loss of charged particles moving parallel to a solid 
surface (stopping power of the solid), which we obtain from the knowledge of the imaginary 
part of the $GW$ self-energy of many-body theory.~\cite{hedin69} In Sec. IV, we present numerical 
calculations of the surface-response function and the electronic stopping power of a Mg(0001) 
surface. We compare our results with the stopping power obtained with the use 
of either the jellium model or a 1D model potential. 
In Sec. V, our conclusions are presented. Unless otherwise is stated, we use 
atomic units throughout, i.e., $e^{2}=\hbar=m_{e}=1$. 
 
\section{Surface-response function} 
 
Take a system of $N$ interacting electrons of density $n_0({\bf r})$ exposed to a 
frequency-dependent external potential $V^{ext}({\bf r};\omega)$. Keeping terms of first order in 
the external perturbation and neglecting retardation effects, the rate $w(\omega)$ at which the 
frequency-dependent external potential $V^{ext}({\bf r};\omega)$ generates electronic excitations 
in the many-electron system~\cite{liebsch} is found to be given by the following expression: 
\begin{equation} 
w(\omega)=-2\,{\rm Im}\int d{\bf r} \int d{\bf r}' \, V^{ext}\,({\bf r};\omega)\,\chi({\bf r},{\bf r}';\omega)\, 
V^{ext}({\bf r}';\omega), 
\end{equation} 
where $\chi({\bf r},{\bf r'};\omega)$ represents the so-called density-response function of the 
many-electron system. In the framework of time-dependent density-functional theory (TDDFT),~\cite{gross95} 
the {\it exact} density-response function of an interacting many-electron system is found to obey the 
following Dyson-type equation: 
\begin{multline}\label{eq5} 
\chi ({\bf r},{\bf r}';\omega)= 
\chi^{0} ({\bf r},{\bf r}';\omega)+ 
\int d{\bf r}_{1} \int d{\bf r}_{2} 
\chi^{0} ({\bf r},{\bf r}_{1};\omega)  \\ 
\times \left\{v({\bf r}_{1},{\bf r}_{2})+f^{xc}[n_0]({\bf r}_1,{\bf 
r}_2;\omega)\right\} \chi ({\bf r}_{2},{\bf r}';\omega). 
\end{multline} 
Here, $\chi^{0}({\bf r},{\bf r}';\omega)$ is the density-response function of 
noninteracting Kohn-Sham electrons,
$v({\bf r}_{1},{\bf r}_{2})$ is the bare Coulomb interaction,
 and $f^{xc}[n_{0}]({\bf r},{\bf r}';\omega)$ represents the 
so-called exchange-correlation (xc) kernel, which is the second functional derivative of the xc 
energy functional evaluated at the ground-state electron density $n_{0}({\bf r})$. In the RPA, 
$f^{xc}[n_{0}]({\bf r},{\bf r'};\omega)=0$. 
 
In the case of a periodic surface, we introduce 2D Fourier expansions of the form 
\begin{equation}\label{eq7} 
\chi({\bf r},{\bf r}',\omega)=\frac{1}{A} 
\sum_{\bf q}^{SBZ}\sum_{{\bf g},{\bf g}'}e^{i({\bf q}+{\bf g})\cdot 
{\bf r_{\parallel}}}e^{-i({\bf q}+{\bf g'})\cdot{\bf r'_{\parallel}}} 
\chi_{\bf g,g'}(z,z';{\bf q},\omega), 
\end{equation} 
where $A$ is a normalization area, ${\bf q}$ is a 2D wave-vector in the surface Brillouin zone (SBZ), 
and ${\bf g}$ and ${\bf g}'$ denote 2D reciprocal lattice vectors. For an external potential of the form 
\begin{equation} 
V^{ext}({\bf r},\omega)=-{\frac{4\pi^2}{A}}\sum_{\bf q}^{SBZ}\,{\rm e}^{i{\bf q}\cdot({\bf r}-iz)}/q, 
\end{equation} 
the rate $w(\omega)$ takes the form~\cite{review} 
\begin{equation} 
w(\omega)=\sum_{\bf q}^{SBZ}w({\bf q},\omega), 
\end{equation} 
where $w({\bf q},\omega)$ denotes the rate at which the external potential generates electronic excitations 
of frequency $\omega$ and 2D wave vector ${\bf q}$: 
\begin{equation}\label{eloss} 
w({\bf q},\omega)={\frac {4\pi}{2A}}\,{\rm Im}g({\bf q},\omega), 
\end{equation} 
with 
\begin{equation}\label{sres} 
g({\bf q},\omega)=-{\frac{2\pi}{q}}\int dz\int dz'\,{\rm e}^{q(z+z')}\, 
\chi_{{\bf g}=0,{\bf g}'=0}(z,z';{\bf q},\omega). 
\end{equation} 
In the RPA, the Fourier coefficients
$\chi_{{\bf g},{\bf g}'}(z,z';{\bf q},\omega)$ are found to 
obey the following matrix equation: 
\begin{multline}\label{eq8} 
\chi_{\bf g,g'}(z,z';{\bf q},\omega)=\chi^{0}_{\bf g,g'}(z,z';{\bf q},\omega) 
\\ 
+\sum_{{\bf g}_1}\int dz_{1} \int dz_{2} 
\,\chi^{0}_{{\bf g},{\bf g}_1}(z,z_1;{\bf q},\omega) 
\\ 
 \times 
v_{{\bf g}_1}(z_1,z_2;{\bf q}) 
\,\chi_{{\bf g}_1,{\bf g}'}(z_{2},z';{\bf q},\omega), 
\end{multline} 
where $v_{\bf g}(z,z';{\bf q})$ denote the 2D Fourier coefficients of the bare 
Coulomb interaction $v({\bf r},{\bf r'})$, 
\begin{equation}\label{eq9} 
v_{\bf g}(z,z';{\bf q})=\frac{2\pi}{|{\bf q}+{\bf g}|}e^{-|{\bf q}+{\bf g}||z-z'|}, 
\end{equation} 
and $\chi^{0}_{\bf g,g'}(z,z';{\bf q},\omega)$ represent the Fourier coefficients of the density-response function of noninteracting Kohn-Sham electrons. 

For positive frequencies, the imaginary part of the Fourier coefficients
$\chi^{0}_{\bf g,g'}(z,z';{\bf q},\omega)$ is easily obtained from the spectral function $S^{0}_{\bf g,g'}(z,z';{\bf q},\omega)$, as follows
\begin{equation}\label{chi0-s0} 
{\rm Im}\chi^0_{{\bf g},{\bf g}'}(z,z';{\bf q},\omega) = -\pi 
S^0_{{\bf g},{\bf g}'}(z,z';{\bf q},\omega), 
\end{equation} 
where 
\begin{widetext} 
\begin{eqnarray}\label{s0} 
S^0_{{\bf g},{\bf g}'}(z,z';{\bf q},\omega)&=& \frac{2}{A} \sum_{\bf 
k}^{SBZ}\sum_{n}^{occ}\sum_{n'}^{unocc} 
 \langle \psi_{{\bf k},n}({\bf r})
|e^{-i({\bf q}+{\bf g}){\bf r}_{\parallel}}| \psi_{{\bf k}+{\bf q},n'} ({\bf 
r})\rangle \langle \psi_{{\bf k}+{\bf q},n'}({\bf r}')|e^{i({\bf 
q}+{\bf g}'){\bf r_{\parallel}'}}| \psi_{{\bf k},n}({\bf r}') 
\rangle \nonumber\\ 
&\times&\delta(\varepsilon_{{\bf k},n}-\varepsilon_{{{\bf k}+{\bf 
q}},n'}+\omega). 
\end{eqnarray} 
\end{widetext}
Here, the sum over $n$ and $n'$ run over all occupied and unoccupied bands,
respectively, and $\varepsilon_{{\bf k},n}$ and $\psi_{{\bf k},n}(z)$ represent,
respectively, the single-particle energies and wave functions of a Kohn-Sham Hamiltonian with an effective potential that is periodic in the plane of the surface. For the evaluation of the real part of
$\chi^0_{{\bf g},{\bf g}'}(z,z',{\bf q},\omega)$, we perform a Hilbert transform of the corresponding imaginary part.\cite{sipipssa08}

The function $g({\bf q},\omega)$ of Eq.~(\ref{sres}) is the so-called surface-response function, which 
represents a key quantity in the description of both single-particle 
and collective excitations at solid surfaces, and whose imaginary 
part [see Eq.~(\ref{eloss})] yields the rate at which an external 
potential generates electronic excitations. Equation~(\ref{sres}) 
shows that only the diagonal Fourier coefficient $\chi_{\bf 
g,g}(z,z';{\bf q},\omega)$ with ${\bf g}=0$ enters the expression 
for the surface-response function. Nevertheless, the full matrix 
nature of $\chi_{\bf g,g'}^0(z,z';{\bf q},\omega)$ and $\chi_{\bf 
g,g'}(z,z';{\bf q},\omega)$ is involved when solving 
Eq.~(\ref{eq8}). These are the so-called local-field effects,\cite{adler} 
which typically play an important role in the presence of strong 
electron-density inhomogeneities but are found to be negligible in 
the case of simple metals like Mg.\cite{silkin} 
 
\section{Electronic stopping power} 
 
Let us consider a probe particle of charge $Z_1$ moving in an inhomogeneous many-electron 
system. The decay rate $\tau_i^{-1}$ of the particle in the state $\phi_{i}({\bf r})$ 
with energy $\varepsilon_{i}$ is obtained from the knowledge of the imaginary part of the self-energy 
$\Sigma({\bf r},{\bf r'};\varepsilon_{i})$, according to~\cite{campillo0} 
\begin{equation}\label{eq12} 
\tau_i^{-1}=-2 \int d{\bf r}\int d{\bf r'} \phi_{i}^{*}({\bf r}){\rm 
Im}\Sigma({\bf r},{\bf r'};\varepsilon_{i})\phi_{i}({\bf r'}). 
\end{equation} 
 
In the $GW$ approximation of many-body theory,\cite{hedin69} and replacing the 
probe-particle Green function by that of a noninteracting particle, 
one finds: 
\begin{multline}\label{eq13} 
\tau_i^{-1}=-2Z_{1}^{2}\int d{\bf r}\int d{\bf r'}\phi_{i}^{*}({\bf r}) 
\\ 
\times\sum_{f}\phi_{f}^{*}({\bf r'}){\rm Im}W({\bf r},{\bf r'}; 
\varepsilon_{i}-\varepsilon_{f})\phi_{f}({\bf r})\phi_{i}({\bf r'}), 
\end{multline} 
where the sum is extended over a complete set of final states $\phi_{f}({\bf 
r})$ of energy $\varepsilon_{f}$. $W({\bf r},{\bf r'};\varepsilon_{i}-\varepsilon_{f})$ 
is the screened interaction of the system, 
which is related to the interacting density-response function as follows 
\begin{multline}\label{eq14} 
W({\bf r},{\bf r'};\omega)=v({\bf r},{\bf r'}) \\ 
+\int d{\bf r}_{1} 
\int 
d {\bf r}_{2} v({\bf r},{\bf r}_{1})\chi({\bf r}_{1},{\bf r}_{2};\omega)v({\bf 
r}_{2},{\bf r'}). 
\end{multline} 
 
In the case of a recoilless point particle moving at a given impact vector 
${\bf b}$ with nonrelativistic velocity ${\bf v}$, the probe-particle initial 
and final states can be described by plane waves in the direction of motion 
and a Dirac $\delta$ function in the transverse direction, i.e., 
\begin{equation}\label{eq15} 
\phi({\bf r})=\frac{1}{\sqrt{A}} e^{i{\bf v}\cdot{\bf r}}\sqrt{\delta({\bf 
r_{\perp}}-{\bf b})}, 
\end{equation} 
where ${\bf r_{\perp}}$ represents the position vector perpendicular to the 
projectile velocity. One then finds that the decay rate of Eq.~(\ref{eq13}) 
can be written as follows 
\begin{equation}\label{eq16} 
\tau_i^{-1}=\frac{1}{\rm T}\sum_{\bf q}P_{\bf q}, 
\end{equation} 
where $T$ is a normalization time and $P_{\bf q}$ is given by the following expression: 
\begin{multline}\label{eq17} 
P_{\bf q}=\frac{4\pi}{\Omega}Z_{1}^{2}\int^{\infty}_{0}d\omega 
\int \frac{d{\bf q}'}{2\pi^{3}}e^{i{\bf b}\cdot({\bf q}+{\bf q}')}  \\ 
\times{\rm Im}W({\bf q},{\bf q}',\omega)\delta(\omega-{\bf q}\cdot{\bf v}) 
\delta(\omega+{\bf q}'\cdot{\bf v}). 
\end{multline} 
Here, $\Omega$ is a normalization volume and $W({\bf q},{\bf q}';\omega)$ represents the double 
Fourier transform of the screened interaction $W({\bf r},{\bf r}';\omega)$. 
 
The quantity $P_{\bf q}$ entering Eq.~(\ref{eq16}) can be interpreted as the probability for the 
probe particle to transfer momentum ${\bf q}$ to the many-electron system. Hence, the stopping 
power of the many-electron system, i.e, the average energy lost by the particle per unit path 
length is found to be given by the following expression: 
\begin{equation}\label{eq18} 
-\frac{dE}{dx}={\frac{1}{L}}\,\sum_{\bf q}({\bf q}\cdot {\bf v})P_{\bf q}, 
\end{equation} 
where $L$ is a normalization length and ${\bf q}\cdot{\bf v}$ represents the energy transfer 
associated to the momentum transfer ${\bf q}$. 
 
Now we restrict our attention to the case of charged recoilless particles moving with 
constant velocity ${\bf v}$ along a definite trajectory at a fixed distance $z$ from a periodic 
solid surface. If one introduces 2D Fourier expansions of the form of Eq.~(\ref{eq7}), 
then Eqs.~(\ref{eq17}) and (\ref{eq18}) yield the following expression for the stopping power: 
\begin{multline}\label{eq27} 
-\frac{dE}{dx}(z)=-\frac{2Z_{1}^{2}}{vA}\sum_{{\bf g},{\bf K}}\sum_{\bf q}^{SBZ} 
e^{i{\bf K}\cdot{\bf b}}\,{\bf q}\cdot{\bf v}\, 
\\ 
\times {\rm Im}W_{{\bf g},{\bf g}+{\bf K}}(z,z;{\bf q},{\bf q}\cdot{\bf v}), 
\end{multline} 
the sum over ${\bf K}$ being restricted to those reciprocal lattice vectors that 
are perpendicular to the velocity of the projectile, i.e., ${\bf K}\cdot{\bf v}=0$. 
 
At this point, we focus on the special situation where the coordinate $z$ is located far 
from the surface into the vacuum. Equation~(\ref{eq14}) shows that under such conditions 
the Fourier coefficients $W_{{\bf g},{\bf g}'}({\bf q},\omega)$ take the following form: 
\begin{multline}\label{eq25} 
W_{\bf g,g'}(z,z;{\bf q},\omega)= 
v_{\bf g}(z,z,{\bf q})\delta_{\bf g,g'} 
 \\ 
-\frac{2\pi q}{|{\bf q}+{\bf g}||{\bf q}+{\bf g'}|} 
g_{\bf g,g'}({\bf q},\omega)e^{-(|{\bf q}+{\bf g}|+|{\bf q}+{\bf g'}|)z}, 
\end{multline} 
where 
\begin{multline}\label{sres2} 
g_{\bf g,g'}({\bf q},\omega)=-{\frac{2\pi}{q}} 
\\ 
\times 
\int dz\int dz'\,{\rm e}^{|{\bf q}+{\bf g}|z}\, 
\chi_{{\bf g},{\bf g}'}(z,z';{\bf q},\omega) {\rm e}^{|{\bf q}+{\bf g'}|z'}, 
\end{multline} 
which for ${\bf g}={\bf g}'=0$ yields the surface-response function entering Eq.~(\ref{eloss}). 
 
The symmetry of the one-particle Bloch states results in the following 
identity: 
\begin{equation}\label{eq28} 
g_{\bf g,g'}(S{\bf q},\omega)=g_{S^{-1}{\bf g},S^{-1}{\bf g'}}({\bf q},\omega), 
\end{equation} 
with $S$ representing a point group symmetry operation in the periodic crystal. 
As a consequence, the stopping power of Eq. (\ref{eq27}) can be evaluated from 
the knowledge of the screened interaction corresponding to wave vectors lying in 
the irreducible element of the surface Brillouin zone (ISBZ). If crystal local-field 
effects are neglected altogether, introducing Eq.~(\ref{sres2}) into Eq.~(\ref{eq27}) yields 
\begin{multline}\label{eq29} 
-\frac{dE}{dx}(z)=\frac{2Z_{1}^{2}}{vA}\sum_{\bf q}^{ISBZ}\sum_{S} 
 \\ 
\frac{2\pi}{|S{\bf q}|}(S{\bf q}\cdot{\bf v}) 
e^{-2|S{\bf q}|z}{\rm Im}\,g({S\bf q},{\bf q}\cdot{\bf v}), 
\end{multline} 
where $g({\bf q},\omega)$ represents the surface-response function 
of Eq.~(\ref{sres}). 
 
In the simplest possible model of a solid surface, in which a semi-infinite 
electron gas described by a Drude dielectric function 
$\epsilon(\omega)=1-\omega_p^{2}/\omega^{2}$ is separated by a planar interface 
at $z=0$ from a semi-infinite vacuum, both Eqs. (\ref{eq27}) and (\ref{eq29}) reduce, for particle 
trajectories outside the solid ($z>0$), to the classical expression:~\cite{ep0} 
\begin{equation}\label{eq30} 
-\frac{dE}{dx}(z)=Z_{1}^{2}\,\frac{\omega_{s}^{2}}{v^{2}}\,K_{0}(2\omega_{s}z/v), 
\end{equation} 
where $K_{0}(x)$ in the zero-order modified Bessel function,~\cite{abra} and $\omega_{s}$ 
is the SP frequency $\omega_{s}=\omega_{p}/\sqrt{2}$. This expression, which 
is known to hold for high particle velocities ($v>>v_{F}$) at large distances 
from the surface ($z>>\lambda_{F}$) shows that under these conditions the 
energy loss is dominated by the excitation of surface plasmons. 
 
\section{Results} 
 
Magnesium ($1s^{2}2s^{2}2p^{6}3s^{2}$) is a monoatomic solid with 
the hexagonal close-packed crystal structure. The input of our 
parameter-free first-principles stopping-power calculations is the 
surface-response function $g({\bf q},\omega)$ of Eq.~(\ref{sres}), 
which we have calculated for the (0001) surface of Mg. The results 
presented below have been found to be well converged for all 
velocities under study. 
The single-particle Kohn-Sham orbitals 
$\psi_{{\bf k},n}({\bf r})$ entering Eq.~(\ref{s0}) were expanded in 
a plane-wave basis set with a kinetic-energy cutoff of 13 Ry. In the 
Fourier expansion of the noninteracting density-response matrix 
$\chi^0_{{\bf g},{\bf g}'}(g_z,g'_z,{\bf q},\omega)$ we included 
$101$~$g_z$-vectors and the components with ${\bf g}={\bf g}'=0$ 
only because of the negligible local-field effects along the 
surface.\cite{silkin} In Eq. (\ref{s0}) all occupied and unoccupied 
energy bands up to 50 eV above the Fermi level were taken into 
account. Numerically, in the evaluation of $S^0_{{\bf g},{\bf 
g}'}(g_z,g'_z,{\bf q},\omega)$ the $\delta$-function was replaced by 
a Gaussian with a broadening parameter $\sigma=0.1$ eV. The sampling 
of the SBZ required for the evaluation of both the surface-response 
function of Eq.~(\ref{sres}) and the stopping power of 
Eq.~(\ref{eq29}) has been performed including 7812 mesh points in 
the SBZ. We set $Z_{1}=\pm 1$, but our results can be used for 
arbitrary values of $Z_{1}$, as the stopping power is, within 
linear-response theory, proportional to $Z_{1}^{2}$. 
 
We compare our first-principles calculations with the results that we have also obtained 
by replacing the actual (0001) surface of Mg by (i) a jellium surface with an 
electron-density parameter $r_s=2.66$ (corresponding to an electron density equal 
to that of valence electrons in Mg) and (ii) a model surface described by the 1D 
potential of Ref.~\onlinecite{echenique97}. This potential 
describes the main features of the surface electronic structure of Mg: the energy 
gap and the Shockley surface state at the $\bar\Gamma$ point (${\bf k}=0$) of the SBZ; in this case, 
we have used films of up to 41 layers of atoms with a lattice parameter $a=4.923\,{\rm a.u.}$ 
corresponding to a film thickness of $100.92\,{\rm a.u.}$, and the 
work function has been taken to be $\Phi=3.66$ eV. 
 
Our first-principles calculations employ a supercell geometry with slabs 
containing 16 atomic layers of Mg that are separated by vacuum intervals. 
The slab geometry imposes a low limit for the 
momentum ${\bf q}$ below which the SP splits into two slab excitations of the form\cite{liebsch} 
\begin{equation} 
w_{\pm}({\bf q})=w_{sp}(1\mp e^{-{q}L})^{1/2}, 
\end{equation} 
with $L$ representing here the slab thickness. This drawback can be softened by increasing the 
slab thickness. 
 
Figure~\ref{img} shows the self-consistent calculations of the 
imaginary part of the surface-response function, ${\rm Im}g({\bf 
q},\omega)$, that we have obtained from Eq.~(\ref{sres}) in the RPA 
for (i) a semi-infinite jellium surface (dashed lines), (ii) the 1D 
model surface potential of Ref.~\onlinecite{echenique97} (thin solid 
lines), and (iii) the actual (0001) surface of Mg (thick solid 
lines). For the low 2D wave vectors ${\bf q}$ under study, the 
energy-loss spectra are clearly dominated by a SP contribution at 
$w_{s}\sim 8\,{\rm eV}$, which seems to first shift to lower 
frequencies, as $q$ increases, and then, from $q \sim 0.02$ on, 
towards higher frequencies. This figure shows that for the small 
values of $q$ considered here both jellium and 1D model calculations 
(dashed and thin solid curves) overestimate the SP energy; our 
calculations show, however, that for larger values of the 2D wave 
vector these simplified models predict accurate values of the SP 
energy. 
 
The important message of Fig.~\ref{img} is that at small values of 
$q$ the actual linewidth of the SP is considerably larger than that 
obtained with the use of 1D jellium-like models. This important 
effect is expected to impact considerably the actual stopping power 
of the solid surface, especially at large distances from the surface 
where the energy loss is dominated by the excitation of surface 
plasmons associated to very low wave vectors. 
 
\begin{widetext} 
\begin{center} 
\begin{figure}[h!] 
\includegraphics*[scale=0.5]{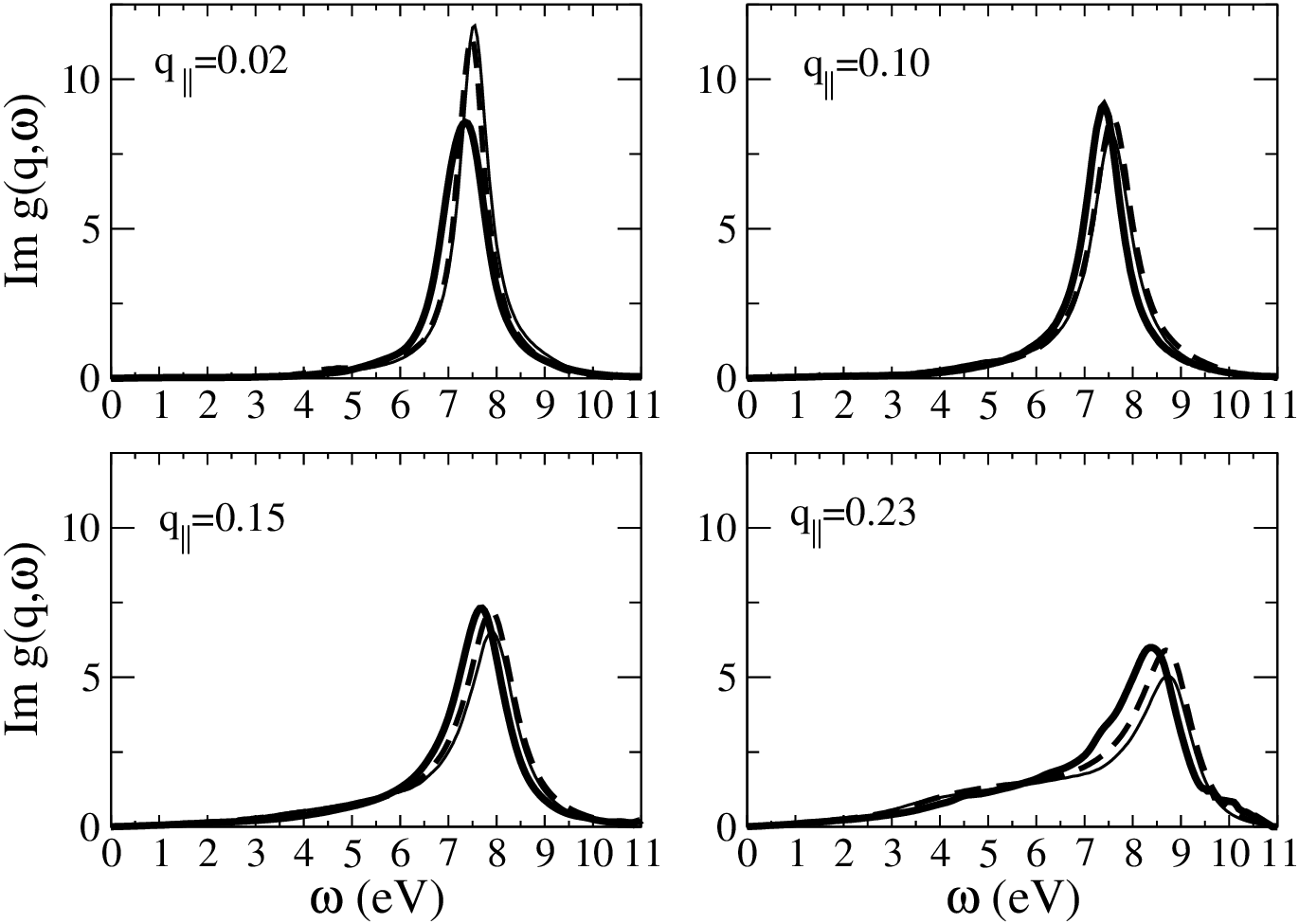} 
\caption{Imaginary part of the RPA surface-response function of 
Mg(0001), ${\rm Im}g({\bf q_{\parallel}},\omega)$, as a function of 
$\omega$, for various values of the magnitude $q$ of the wave 
vector. The thick solid line represents self-consistent {\it 
first-principles} calculations, the dashed line represents the 
corresponding results obtained for a semi-infinite jellium surface, 
and the thin solid line represents the corresponding results 
obtained by using the 1D model potential of 
Ref.~\onlinecite{echenique97}.} \label{img} 
\end{figure} 
\end{center} 
\end{widetext} 
 
\begin{widetext} 
\begin{center} 
\begin{figure} [ht!] 
\includegraphics*[scale=0.5]{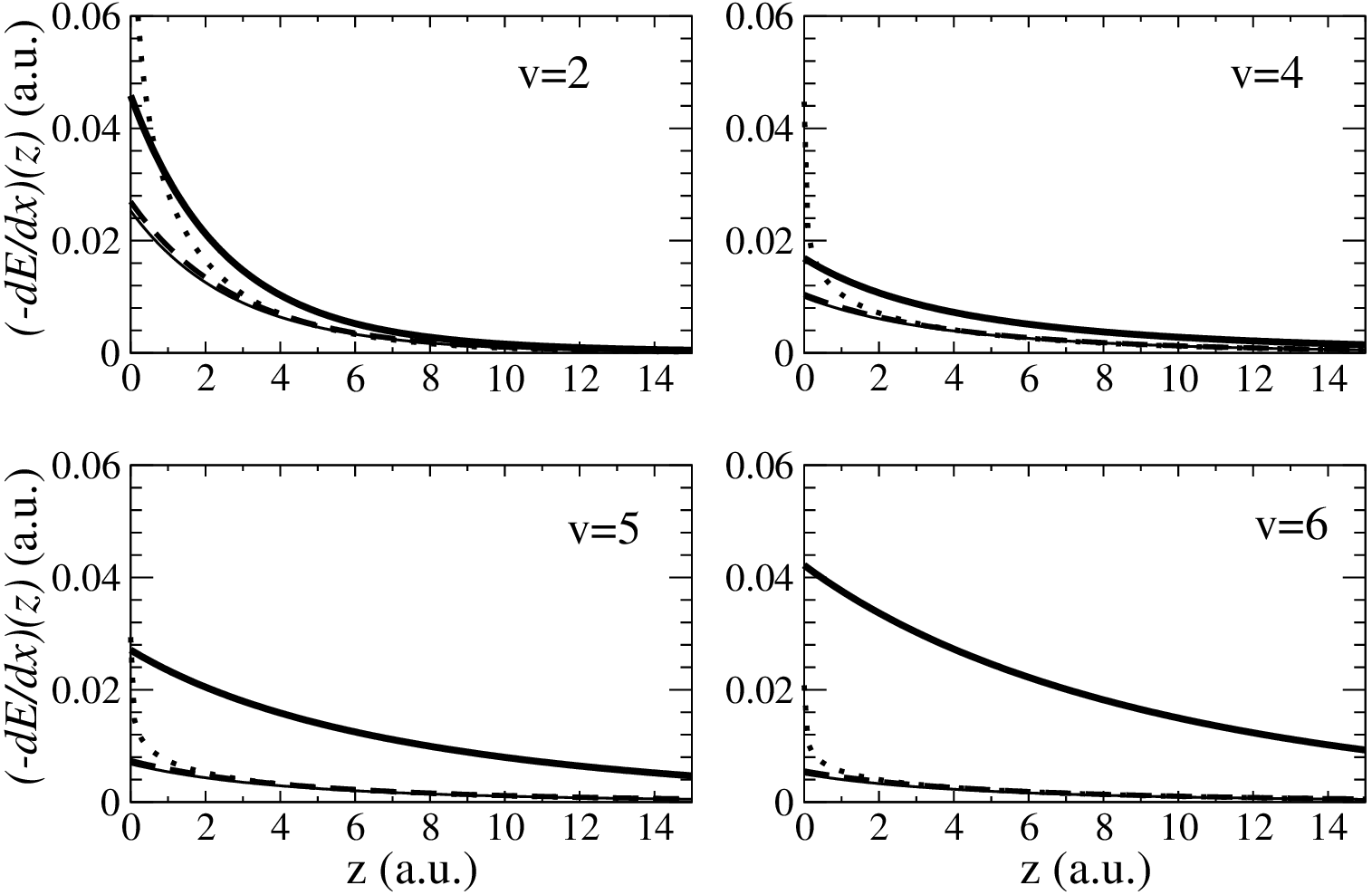} 
\caption{Stopping power of Mg(0001), versus $z$, for a recoilless particle of 
charge $Z_1=1$ moving with various velocities. The thick solid line represents 
self-consistent {\it first-principles} calculations. The dashed and thin solid 
lines represent the corresponding results obtained for a semi-infinite jellium 
surface and by using the 1D model-potential of Ref.~\onlinecite{echenique97}, 
respectively. The dotted lines represent the classical energy loss of Eq.~(\ref{eq30}).} 
\label{eloss-b} 
\end{figure} 
\end{center} 
\end{widetext} 
 
Now we focus on the special situation where an external recoilless 
particle of charge $Z_1=1$ is moving with constant velocity along a 
definite trajectory at a fixed distance $z$ far from the surface 
into the vacuum. Fig.~\ref{eloss-b} exhibits the stopping power of 
Mg(0001) for this moving particle, as obtained from Eq.~(\ref{eq29}) 
in the RPA for (i) a semi-infinite jellium surface (dashed lines), 
(ii) the 1D model surface potential of Ref.~\onlinecite{echenique97} 
(thin solid lines), and (iii) the actual (0001) surface of Mg (thick 
solid lines). At the velocities under consideration ($v>v_F$), the 
energy-loss spectrum of charged particles moving far from the 
surface into the vacuum is dominated by long-wavelength surface 
excitations (small $q$), i.e., by the excitation of surface 
plasmons. Hence, we might be tempted to conclude that 
Eq.~(\ref{eq30}) (represented in Fig.~\ref{eloss-b} by a dotted 
line) should be a good representation of the actual stopping at 
$z>>\lambda_{F}$. Indeed, Fig.~\ref{eloss-b} shows that this 
classical limit is in excellent agreement at large values of $z$ 
with the results obtained with the use of 1D jellium-like models 
(dashed and thin solid lines). However, it is important to note that 
these models do not account for the intrinsic linewidth of surface 
plasmons which, as a result of interband transitions that are absent 
in these simplified models, dominates the energy loss at large 
distances. As the velocity increases (see the lower panels of 
Fig.~\ref{eloss-b}), lower  values of the wave vector enter the 
excitation spectrum leading to an increased impact of the intrinsic 
surface-plasmon linewidth on the stopping power and, therefore, more 
pronounced differences between the stopping power of a jellium-like 
surface (dashed and thin solid lines) and the real surface (thick 
solid lines). 
 
\begin{figure} 
\includegraphics*[scale=0.3]{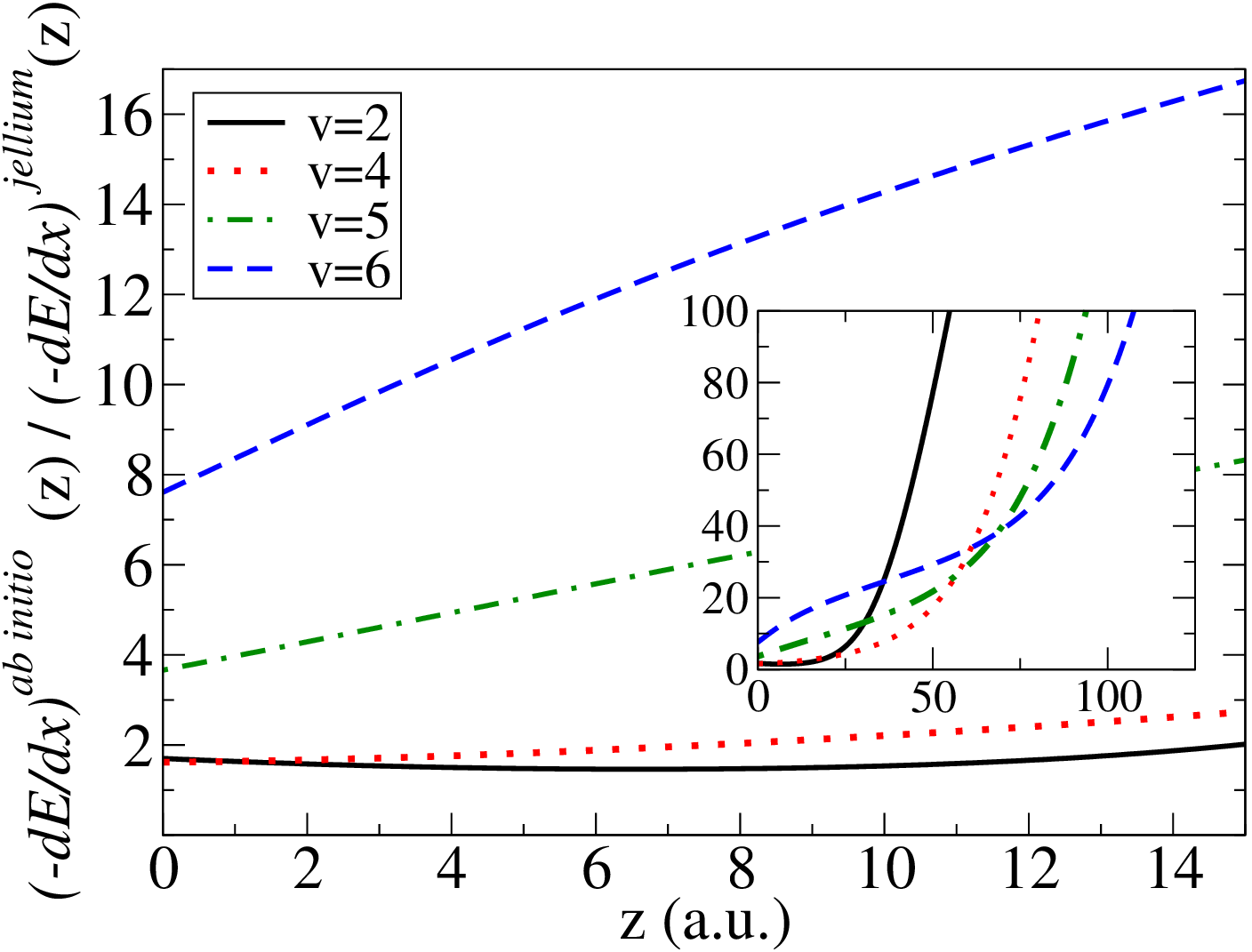} 
\caption{ (Color online) Ratio $(-dE/dx)^{\it ab \, 
initio}(z)/(-dE/dx)^{jellium}(z)$, versus $z$, for a recoilless 
particle of charge $Z_1=1$ moving with various velocities outside 
theMg(0001) surface. In the inset, larger distances $z$ from the 
surface are considered.} \label{ratio} 
\end{figure} 
 
Figure~\ref{ratio} shows the ratio between the {\it first-principles} stopping power 
of the real Mg(0001) surface and the stopping power of the corresponding jellium surface 
[which at large distances coincides with the classical result of Eq.~(\ref{eq30})] 
for the four values of the velocity considered in Fig.~\ref{eloss-b}. 
The {\it ab initio} energy loss is considerably larger than that obtained from the 
classical Eq.~(\ref{eq30}) at all large distances from the metal surface. 
As $z\to\infty$, the classical Eq.~(\ref{eq30}) (which assumes the linewidth of the 
surface plasmon to be zero) decays exponentially; indeed, at these large distances from 
the surface the stopping power is dominated by the finite intrinsic linewidth of the 
surface plasmon, leading to a ratio that increases exponentially. This exponential 
increase also occurs when the stopping power of the real surface is replaced by that 
of a semi-infinite electron gas described by a Drude dielectric function of the form 
$\epsilon(\omega)=1-\omega_p^2/\omega(\omega+i\gamma)$, $\gamma$ being a damping parameter 
that would give account approximately for the finite surface-plasmon linewidth.~\cite{nunez} 
 
\section{Summary and conclusions} 
\label{sec:C} 
 
We have carried out first-principles self-consistent calculations of the 
surface-response matrix and the stopping power of the (0001) surface of Mg. 
Our results indicate that band-structure effects (and, in particular, interband transitions) 
play a key role in the description of the asymptotic behaviour of the stopping power far from 
the surface, even in the case of a free-electron-like metal such us Mg. In particular, we find 
that the linewidth of the surface-response function is considerably enhanced at small wave 
vectors, which yields an increased energy loss of charged particles moving far from the 
surface that cannot be described by simplified 1D jellium-like models. This important effect, 
which should be present in the case of all metal surfaces, is expected to be relevant for the 
understanding of the interaction between charged particles and the internal surface of microcapillaries. 
New experiments along these lines would be desirable.
 
\section{acknowledgments} 
The authors acknowledge partial support by the UPV/EHU 
(GIC07IT36607), the Basque Unibertsitate eta Ikerketa Saila, the 
Spanish Ministerio de Educaci\'{o}n y Ciencia (Grants No. CSD2006-53 
and FIS2007-066711-CO2-00), and the EC 6th framework Network of 
Excellence NANOQUANTA. The work of V.M.S. is sponsored by the 
IKERBASQUE FOUNDATION.

\end{document}